\documentclass[a4paper,conference]{IEEEtran}
\usepackage{times}
\usepackage{balance,amsmath,graphicx,amssymb,cite}
\usepackage{booktabs}
\usepackage[pagebackref,breaklinks,colorlinks]{hyperref}
\usepackage{pifont}
\newcommand{\cmark}{\ding{51}}%
\newcommand{\xmark}{\ding{55}}%
\newcommand{\ra}[1]{\renewcommand{\arraystretch}{#1}}

\begin{document}

\title{Attention W-Net: Improved Skip Connections for Better Representations}
\author{
\IEEEauthorblockN{Shikhar Mohan$^*$, Saumik Bhattacharya$^*$ and Sayantari Ghosh$^\dagger$ \\
$^*$ Indian Institute of Technology Kharagpur, India\\
$^\dagger$ National Institute of Technology Durgapur, India}}

\maketitle


\begin{abstract}
Segmentation of macro and microvascular structures in fundoscopic retinal images plays a crucial role in the detection of multiple retinal and systemic diseases, yet it is a difficult problem to solve. Most neural network approaches face several issues such as lack of enough parameters, overfitting and/or incompatibility between internal feature-spaces. We propose Attention W-Net, a new U-Net based architecture for retinal vessel segmentation to address these problems. In this architecture, we have two main contributions: Attention Block and regularisation measures. Our Attention Block uses attention between encoder and decoder features, resulting in higher compatibility upon addition. Our regularisation measures include augmentation and modifications to the ResNet Block used, which greatly prevent overfitting. We observe an F1 and AUC of 0.8407 and 0.9833 on the DRIVE and 0.8174 and 0.9865 respectively on the CHASE-DB1 datasets — a sizeable improvement over its backbone as well as competitive performance among contemporary state-of-the-art methods.
\end{abstract}

\begin{figure*}[ht]
    \begin{minipage}[b]{1.0\linewidth}
      \centering
      \centerline{\includegraphics[width=14cm]{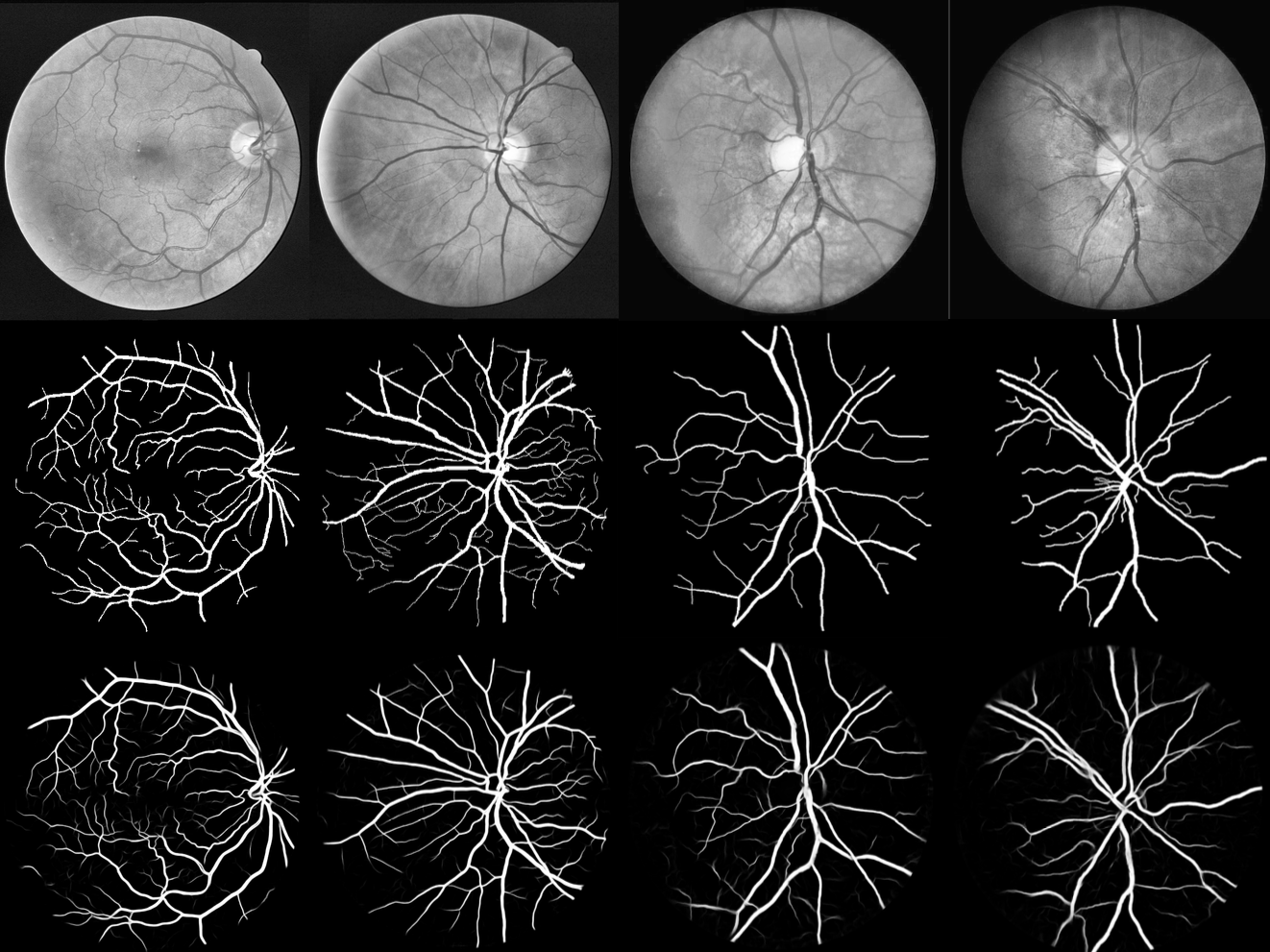}}
      \centerline{}\medskip
    \end{minipage}
    \caption{First two columns are from DRIVE and the second two are from CHASE-DB1. For rows from top to bottom, the labels are (a) Original Image (b) Ground Truth (c) Attention W-Net (ours) Prediction}
    \label{fig:viz}
\end{figure*}

\section{Introduction}
\begin{figure*}[ht]
    \begin{minipage}[b]{1.0\linewidth}
      \centering
      \centerline{\includegraphics[width=15cm]{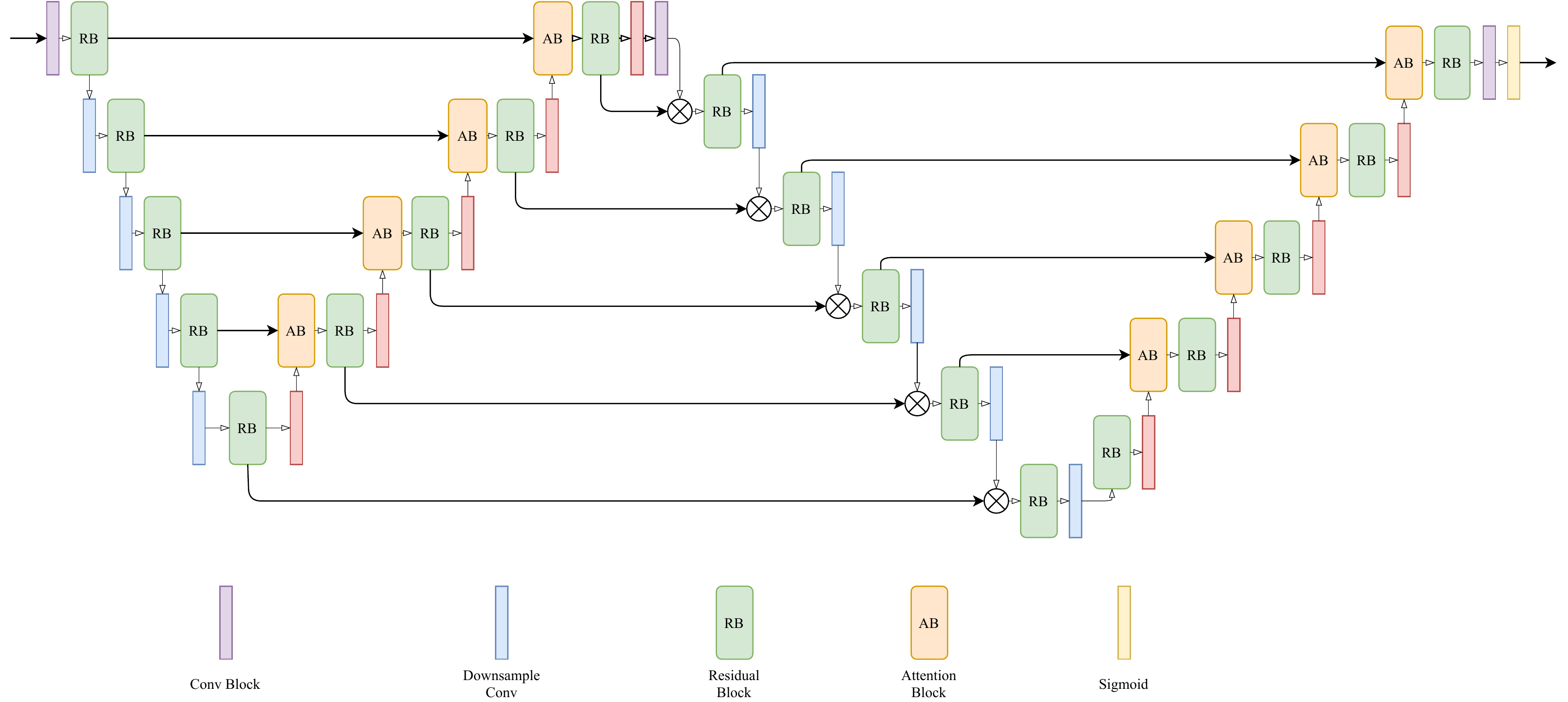}}
      \centerline{}\medskip
    \end{minipage}
    \caption{Attention W-Net Network Structure}
    \label{fig:attnresd}
\end{figure*}
\label{sec:intro}
Learned medical image segmentation is a research area that has been receiving a great deal of attention in the
computer vision community. Traditionally, medical experts have been analysing these various forms of medical images 
ranging from X-Ray, Ultrasound Imaging, Computed Tomography (CT), Magnetic Resonance Imaging (MRI), Fundoscopic 
Imaging, etc. to manually generate dense labels. The above being a tedious, error-prone and expensive task, a
computer-aided diagnosis is a natural solution. Deep learning based methods, more specifically CNN based architectures, have come very far in solving these problems. Many architectures such as FCN \cite{fcn} and U-Net \cite{ronneberger2015u} have given rise to models that have grown in popularity for such tasks due to their near-radiologist performances and much higher efficiency.
In this paper, we focus on the Retinal Blood Vessel Segmentation task using fundoscopic images. These images serve
as an essential \emph{in vivo} test for detecting retinal diseases as well as systemic diseases (such as high blood pressure, diabetic
retinopathy, and microvascular complications of diabetes). Therefore, accurate segmentation of retina blood vessels is an essential task.

The most significant difficulty in the vessel segmentation task is the insignificant difference between the pixels
belonging to the vessels and the pixels belonging to the background, followed by the thinness of microvessels
at the ends and edges. The true scope of this difficulty can be realised when we consider the noise present
in these images and optic disks (bright spots) present in retinal fundus images. Much of this noise can be
attributed to poor illumination, sensor noise and incorrect filters and angles used in fundus cameras. The limited 
amount of retinal data is a sizeable hurdle as well. In fact, there are less than 25 images available for training 
in publicly available datasets, \emph{i.e.} CHASE-DB1 \cite{chasedb1}, DRIVE \cite{drive} and STARE \cite{stare}.

Deep learning has revolutionized various areas in computer vision and biomedical image understanding is no exception. Neural networks have brought about state-of-the-art performance in multiple challenges, and CNN based architectures have done the same in computer vision. Among CNN based neural networks, U-Net \cite{ronneberger2015u} based architectures have been receiving a lot of attention for biomedical image segmentation tasks due to these networks consisting of an encoder to decoder structure with skip connections which allow for an efficient flow of 
gradients for learning. The encoder captures context information whereas the decoder enables precise localisation based on the task.

Many other research works investigate different aspects of this task, such as \emph{Uysal et. al.} \cite{uysal2021exploring} 
highlights the importance of image augmentation in this task.  \cite{uysal2021exploring} 
investigates not only affine transforms but also transforms in the pixel subspace as well as 
elastic transforms. SGL \cite{zhou2021study} aims to improve the robustness of the model since 
it is effectively trained on noisy labels due to clinicians’ manual labelling. GANs have been 
employed for this task as well, where RV-GAN \cite{kamran2021rv} introduces a feature matching 
loss to ameliorate the type-1 error caused by thresholding segmentation heatmaps to generate 
segmentation maps. Attention based techniques, which find their origin in language processing (NLP) for image captioning \cite{Anderson_2018_CVPR}, machine translation and comprehension tasks \cite{shen2018disan}, have garnered interest for this task as well.
SA-UNet \cite{guo2021sa} introduces a spatial attention module at the bottom-most
layer for adaptive feature refinement. This attention module aims at suppressing unimportant features and
enhancing more important features. Attention U-Net \cite{oktay2018attention}
introduces a modified grid-based attention module at every layer. This form of attention gating filters gradients in the forward as well as the backward pass.

However, most of the above approaches struggle with some common issues. Most of the above networks do not have enough parameters to learn the necessary complex features. This can be observed in that these networks primarily struggle to segment microvessels at the end of larger vessels in the fundus images.
In networks which do have more parameters such as LadderNet  \cite{zhuang2018laddernet} we observe sizeable overfitting, but the detachment from the basic encoder-decoder architecture is a step in the right direction. Even though skip-connections are the reason why U-Net architectures work so well, the difference in feature-spaces between the encoder features and decoder features results in sub-optimal representation learning in all of the aforementioned architectures. Taking all of the above into account, we introduce Attention W-Net. The two key motivators of our work are to help the signal from the encoder branch stay relevant in the decoder branch and to add a meaningful regularisation signal to our training process as well. Hence, the main contributions of our work are as follows:
\begin{itemize}
\item An Attention Block that magnifies the utility of skip connections and ensures learning of richer features by preserving the relevance of the encoder features in the decoding branch.
\item A meaningful regularising signal in the form of careful image augmentations and the construction of a Residual Block to prevent our network from overfitting.
\item A Retina Blood Vessel segmentation model with top performance.
\end{itemize}

\section{Proposed Methodology}

\label{sec:method}

The Attention W-Net employs LadderNet as a backbone network, on top of which we add our 
modified Attention Blocks. The LadderNet framework we employ has two alternating encoder branches and decoder branches, which make the structure akin to two U-Nets. The Residual Block is present at every level, whereas the Attention Block is present only in the two upsampling phases of the network, ensuring smooth integration of the encoder signal from the skip-connections into the decoder layers. In the end, we use Binary Cross-Entropy loss to compute the loss. For pixel $i$ with prediction $\hat{y}_i$ and ground truth $y_i$, we have $L_i$ as the value of the cost function as follows:
\begin{eqnarray}
L_i = -y_i \log (\hat{y}_i) - (1-y_i) \log (1-\hat{y}_i)
\end{eqnarray}

\begin{figure}[htb]
    \begin{minipage}[b]{1.0\linewidth}
      \centering
      \centerline{\includegraphics[width=7cm]{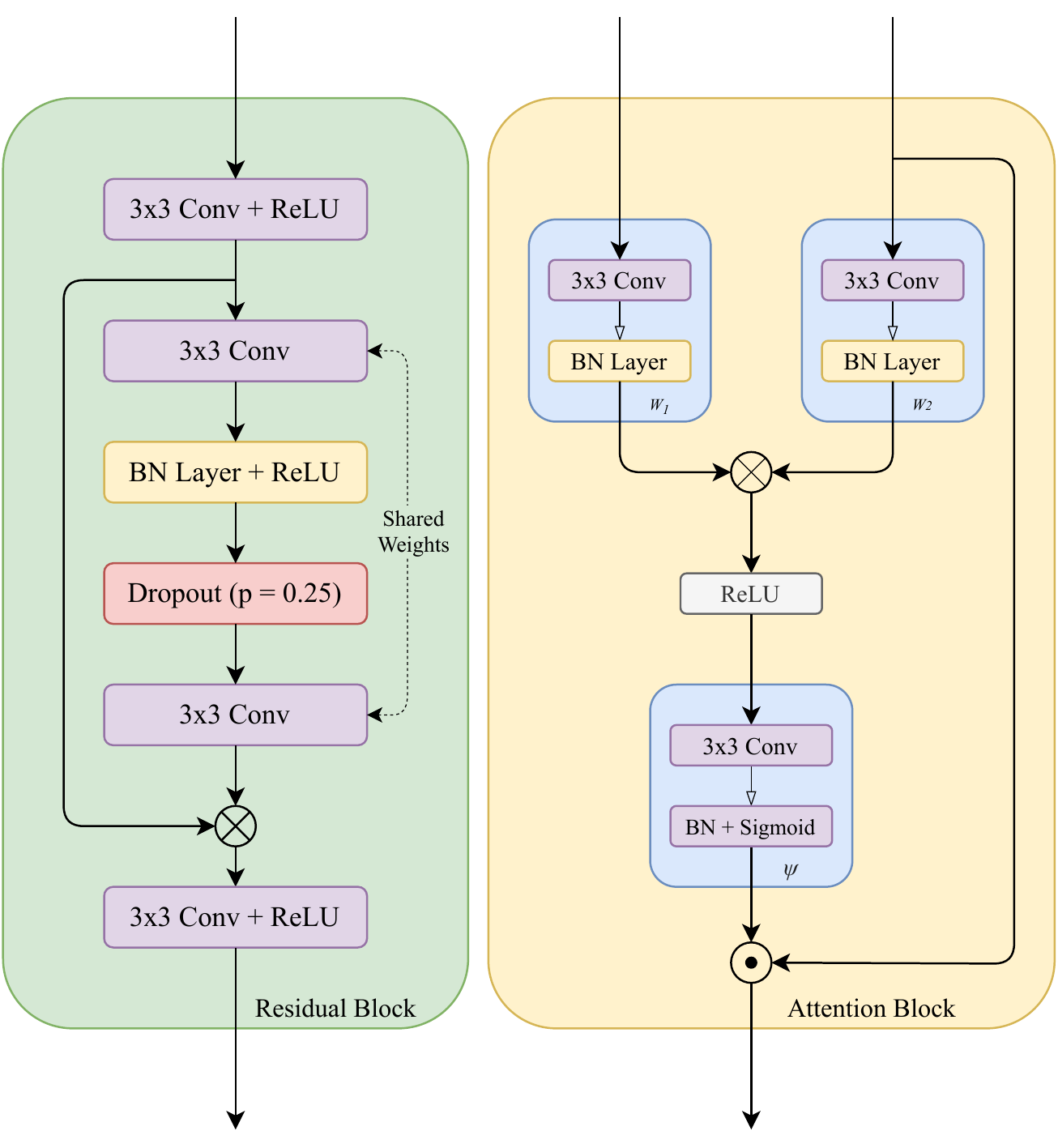}}
    \end{minipage}
    \caption{Residual Block (left) and Attention Block (right).}
    \label{fig:resattn}
\end{figure}
\subsection{Attention Block}
\label{ssec:attn}
In most U-Net based architectures the encoder branch focuses on extracting high-dimensional features from the data, whereas the decoder branch focuses on rearranging and localising these features to achieve the goal at hand -- image segmentation in this case. It has been very well observed that the decoding process benefits from being able to incorporate the encoder signal as well, which is implemented by summing or concatenating the encoder features using skip-connections. However, it is easy to see that these features, if used directly, would be incompatible to some extent due to the innate differences between the two feature spaces at hand.

To tackle this issue, we introduce our Attention Block. The structure is similar to the counterpart found in \cite{oktay2018attention}, except we attend over the encoder features using the decoder features, and add the resultant features for further decoding. The difference between the two is substantial as the former results in the encoder signal being drowned out as it is only used to attend over the decoder signal.
Formally, we state that $g^{k-1} \in \mathbb{R}^{H \times W \times C_k}$ and $x^{k-1} \in \mathbb{R}^{H \times W \times C_k} $ are the encoder and decoder feature outputs respectively from the $(k-1)^\text{th}$ layer, $W_1,W_2$ and $\phi$ are convolutional blocks in the Attention Block. Now we have,
\begin{eqnarray}
p^{k-1} = \psi(\text{ReLU}(W_1(g^{k-1}) + W_2(x^{k-1})))
\end{eqnarray}
Here, $p^{k-1}$ denotes the attention map generated from encoder and decoder features from the $(k-1)^{\text{th}}$ layer. All values in $p^{k-1}$ are in $(0,1)$. This attention map is then applied to the encoder feature as follows:
\begin{eqnarray}
g_d^{k-1} = g^{k-1} \odot p^{k-1}
\end{eqnarray}
Where $g_d^{k-1}$ is the attended encoder feature and $\odot$ denotes element-wise multiplication. Now that $g_d^{k-1}$ contains information regarding the decoder feature in it as well, we have $x_d^{k-1} = g_d^{k-1} + x^{k-1}$ where $x_d^{k-1}$ represents the summed up feature which is then sent to the $k^\text{th}$ layer in the decoder network.

\subsection{Regularisation}
\label{ssec:resblk}
Most U-Net based architectures for this task are prone to overfitting, and to solve this problem we emphasize the importance of not only standard regularising methods but meaningful ones as well. In this section, we discuss two components of our work: the Residual Block and our careful image augmentation. The importance of meaningful regularisation is emphasized in this task more than others due to the sheer unavailability of data as well as the defects found therein.

The Residual Block we use is inspired by the ResNet \cite{he2016deep} Basicblock, with 
minor modifications. Majorly, it uses only two convolutional layers, which share weights (as in LadderNet), except 
with added BatchNorm layers to the official LadderNet implementation. The shared weights between the two convolutional 
layers make for 
easier learning of features, in between which we add a dropout layer to prevent overfitting. This 
Residual Block has a lesser number of parameters than a standard ResNet Basicblock, and the residual connection, 
dropout layer and the BatchNorm layers result in an effective regularisation, which helps us take advantage of the large number of parameters we include in this network to learn richer representations without overfitting.

Note that this doesn't contradict with the idea that our intention here is to make a deeper network. This is because in many cases, the shallower networks do not have enough capacity to directly perform the segmentation task, and they are often combined with a deeper module or a refinement module \cite{zhang2019automatic, zhang2018context}. Thus, it is important to design the network with sufficient depth. This being said, removing the weight sharing in the residual blocks significantly degrades our performance (8-12\% performance drop in terms of F1-score on DRIVE dataset). Thus, the proposed model is designed after performing careful ablation studies keeping a trade-off between the number of parameters and the performance of the model. 

In addition to this, we use careful image augmentation as well in our training process. Table \ref{table:2} details the image augmentation techniques we use. These carefully selected image augmentations make our network robust to the naturally occurring distortions in the dataset as well as increase its statistical size.
\begin{table}[tbp]
\centering
\ra{1.2}
\caption{Random data augmentations used with probabilities}
\label{table:2}
\begin{tabular}{l r}
\toprule
Data Augmentations Used & Probability\\
\midrule
Random Sized Crop with Padding & 0.5 \\
Vertical Flip & 0.5 \\
Rotate 90\textdegree & 0.5 \\ 
Elastic Transform & 0.5 \\
Grid Distortion & 0.5 \\
Optical Distortion & 0.8 \\
Brightness Contrast & 0.8 \\
Random Gamma & 0.8 \\
\bottomrule
\end{tabular}
\end{table}
\section{Experiments}
\label{sec:experiment}
In this section, we provide a detailed description of the experimental design, the results and the comparisons. The experiments have been performed on a Windows workstation, which has a single NVIDIA GeForce RTX 2080 with 8GB VRAM and Intel Core i5-9600K CPU. We use a PyTorch as our framework of choice to design and train our architectures.
\subsection{Data}
\label{ssec:data}
\begin{figure}[ht]
    \begin{minipage}[b]{1.0\linewidth}
      \centering
      \centerline{\includegraphics[width=7.5 cm]{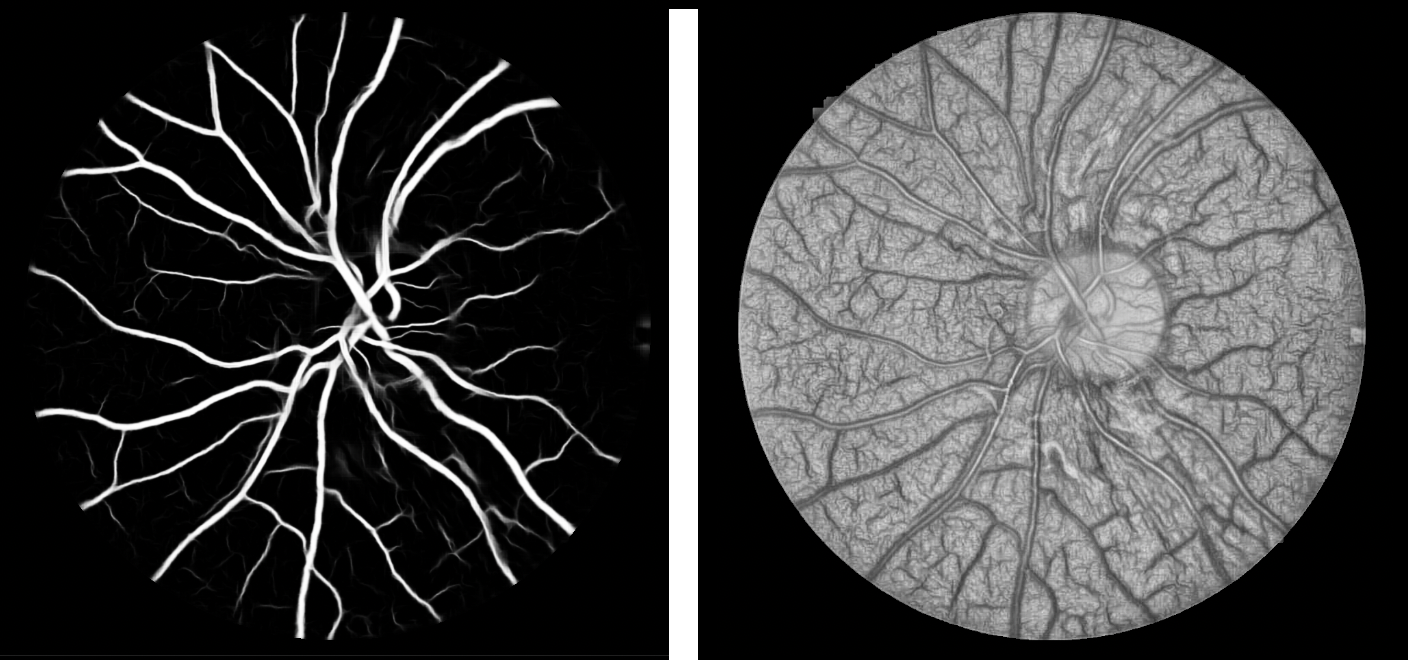}}
    \end{minipage}
    
    \caption{Prediction (left) vs. Seg-Grad-Cam visualisation (right).}
    \label{fig:dataviz}
    
\end{figure}

In the Retina Vessel Segmentation, the most widely used datasets are the DRIVE \cite{drive} and CHASE-DB1 \cite{chasedb1} datasets. The DRIVE dataset contains 20 training and 20 testing images in \emph{.tif}($565 \times 584$), whereas the CHASE-DB1 dataset had 20 training and 8 testing images in \emph{.jpg}($999 \times 960$) format.
For both the datasets, we extract patches of size $48 \times 48$ and use $10\%$ of training samples as validation data. For both inference and training, overlapping patches are extracted, after which we apply the FOV (field of view) masks. The official FOVs are not provided in the case of the CHASE-DB1 dataset, so we extract them from the original images with methods same as in IterNet \cite{li2020iternet}.

\subsection{Training Setup}
\label{ssec:train}
We choose a LadderNet backbone with 5 levels, resulting in 1419636 parameters and the value of dropout was set to 0.25. We evaluate the segmentation loss using binary cross-entropy loss and applied the Adam optimizer with default parameters and a batch size of 1024. We used a ``reduce learning rate on plateau" strategy for learning rate scheduling, where we train with the rate $10^{-3}$ for 100 epochs and $10^{-4}$ for the rest of the training process, which goes on for 250 epochs in total. For the CHASE-DB1 dataset, we train for 50 more epochs with a learning rate $5 \times 10^{-5}$ to compensate for the larger image size leading to a larger training set.

We tried different loss terms, like \textit{l}1 and \textit{l}2 losses, for the segmentation task. However, BCE loss has performed significantly better than the other loss terms. This is in line with the recent SOTA methods \cite{zhuang2018laddernet, li2020joint, su2021dv} where BCE loss and its variants have performed significantly better than the conventional \textit{l}-p norm based losses in segmentation tasks.

We use carefully chosen randomized data augmentation for this training process as well. The data augmentations we used along with probability values can be seen in Table \ref{table:2}. The entire training process takes nearly 15 hours.
\subsection{Inference}
\label{ssec:infer}
During inference, we use a patch-width of 48 and stride length
5 to obtain our results. This setting implies 89.6\% overlap
between the patches. For a pixel, all the
predicted values are averaged over all patches that include
the given pixel to determine the true prediction. This brings better
predictions in our experiments, at the cost of computational
overhead. We observe that averaging produces an ensembling-
like effect and maintains continuity for the thin microvessels
present, which is primarily why higher overlap corresponds to
better performance.

\subsection{Quantitative Bench-marking}
\label{ssec:scores}
We compare our architecture with multiple formative as well as state-of-the-art architectures on the bases of F1-score, AUC--ROC and Accuracy. We perform this comparison over the DRIVE and CHASE-DB1 datasets using their publicly available codebases. Our method might not achieve state-of-the-art performance on these metrics but it is important to note that we base our Attention-Block architecture on top of the LadderNet network, which means the magnitude of improvement in metrics over LadderNet is what we give more importance to. Figure \ref{fig:dataviz} shows raw predictions and (monochrome) attention heatmap for a sample image from the CHASE-DB1 dataset. The heatmap is generated using Seg-Grad-Cam \cite{Vinogradova2020TowardsIS}.

\begin{table}[tbp]
\small
\centering
\ra{1.2}
\caption{Comparison on the DRIVE and CHASE-DB1 datasets.}
\label{table:3}
\begin{tabular}{l r r r}
\toprule
Method  & F1-Score & ACC & AUC\\
\midrule
\emph{\textbf{DRIVE}} \\
DU-Net \cite{jin2019dunet} & 0.8174 & 0.9555 & 0.9752\\
Recurrent UNet \cite{chen2017deeplab}  & 0.8155 & 0.9556 & 0.9782 \\
Residual UNet \cite{chen2017deeplab} & 0.8149 & 0.9553 & 0.9779 \\
R2U-Net \cite{alom2018recurrent}  & 0.7928 & 0.9556 & 0.9634 \\
LadderNet \cite{zhuang2018laddernet}  & 0.8202 & 0.9561 & 0.9793 \\
IterNet \cite{li2020iternet} & 0.8205 & 0.9573 & 0.9816 \\
AG-Net \cite{zhang2019attention}  & - & 0.9692 & 0.9864 \\
SA-UNet \cite{guo2021sa} &  0.8263 & 0.9698 & 0.9864 \\
SGL \cite{zhou2021study}  & 0.8316 & - & 0.9886 \\
\emph{Uysal et. al.} \cite{uysal2021exploring}  & - & 0.9712 & 0.9855 \\
AW-Net (Ours)  & 0.8407 & 0.9588 & 0.9833 \\
\midrule
\emph{\textbf{CHASE-DB1}} \\
DU-Net \cite{jin2019dunet}                               & 0.7853                 & 0.9724       & 0.9863 \\
Residual U-Net \cite{chen2017deeplab}                       & 0.7800                   & 0.9553       & 0.9779 \\
Recurrent U-Net \cite{chen2017deeplab}                      & 0.7810                  & 0.9622       & 0.9803 \\
R2U-Net \cite{alom2018recurrent}                              & 0.7928                 & 0.9634       & 0.9815 \\
LadderNet \cite{zhuang2018laddernet}                           & 0.8031                 & 0.9656       & 0.9839 \\
IterNet \cite{li2020iternet}                            & 0.8073                 & 0.9655       & 0.9851 \\
SA-UNet \cite{guo2021sa}                             & 0.8151                 & 0.9755       & 0.9905 \\
AW-Net (ours)                        & 0.8174                 & 0.9689       & 0.9865 \\
\bottomrule
\end{tabular}
\end{table}

We see that our model has notable gains over the initial LadderNet architecture in all three metrics, which tells us that our novel additions sizeably improve learning. Accuracy being one of the least reliable metrics for image segmentation, we believe it is logical to give more importance to F1-Score and AUC metrics. For most models from IterNet \cite{li2020iternet} onwards (including ours), we note very good performance in terms of AUC ($\geq 0.97$), but we see that there is very little variance. However, there is a significantly higher disparity in F1-Scores, which is where our model excels. It is to be noted that our proposed model outperforms even the most recent approaches. Thus, these benchmarks prove that our novel additions help our backbone capture richer representations, leading to better segmentation performance. 
\subsection{Improvement Analysis}
\label{ssec:stat}
\begin{figure}[tbp]
\centering
    \begin{minipage}[b]{0.8\linewidth}
      \centering
      \centerline{\includegraphics[width=7cm]{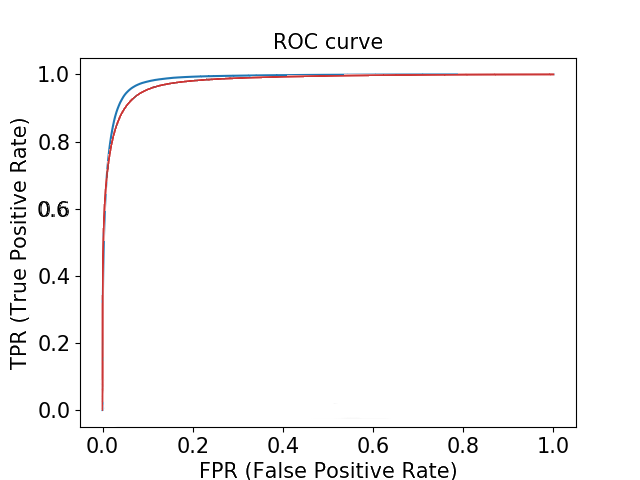}}
    \end{minipage}
    \caption{ROC curves for LadderNet (red) and Attention W-Net (blue).}
    \label{fig:roc}
\end{figure}

In this section, we demonstrate the significance of improvement over the LadderNet backbone. We do this in two ways, first by conducting a statistical significance test over 5 identical runs with only differing randomisation seeds and second by comparing ROC curves, Figure \ref{fig:roc}.
\begin{table}[tbp]
\small
\centering
\caption{Tests for statistical significance }
\label{table:5}
\begin{tabular}{rr|rr}
\toprule
\multicolumn{2}{c}{\textbf{LadderNet}}               & \multicolumn{2}{c}{\textbf{Ours}}                        \\ \hline
\multicolumn{1}{r}{F1-Score} & \multicolumn{1}{r}{AUC} $\vert$& \multicolumn{1}{r}{F1-Score} & \multicolumn{1}{r}{AUC}    \\ \midrule
0.8202                      & 0.9793                  & 0.8407                      & 0.9833                      \\
0.8195                      & 0.9801                  & 0.8390                       & 0.9835                     \\
0.8210                       & 0.9812                  & 0.8374                      & 0.9844                     \\
0.8203                      & 0.9785                   & 0.8397                      & 0.9844                     \\
0.8191                      & 0.9790                  & 0.8389                      & 0.9824                      \\ \bottomrule
\end{tabular}
\end{table}

With 4 degrees of freedom, we conduct a \emph{paired t-Test} with a one-tailed distribution, through which for both F1-Score and AUC, we reject the null hypothesis (with $\alpha = 0.005$). We find p-values for the former to be $5.55 \times 10^{-6}$ and for the latter to be $3.81 \times 10^{-4}$, which confirm that our improvements over LadderNet are statistically significant for both the aforementioned metrics.

\subsection{Ablation Study}
\label{ssec:abla}
In this section, we perform an ablation study, where we make use of the DRIVE dataset and make small changes to the network, keeping the rest of the training procedure invariant. The training procedure we adopt is the same as in Sec. \ref{ssec:train}, i.e. for 250 epochs with ``reduce LR on plateau'' learning rate scheme, starting with 0.01. Note that without any of our novel elements, we essentially have the LadderNet architecture, which is the starting point of our ablation study.

We demonstrate the impact of our three contributions: ResBlock modification, Regularisation and Attention Block (AB). To understand the effects of our proposed blocks, we add them sequentially to the backbone model to investigate the improvements achieved in each step. Additionally, we also show how our design of the Attention Block -- which we refer to as Type-2 -- is superior in performance to the traditional design (as in AG-Net \cite{zhang2019attention}) which we refer to as Type-1. Maintaining the same definitions from Sec. \ref{ssec:attn}, Type-1 AB can be formally characterized by computing $x_d^k$ (the $k^\text{th}$ layer decoder input) differently from Type-2 AB. Instead of (3), we have
$x_d^{k} = p^{k-1} \odot x^{k-1}$.
The calculated $x_d^k$ is then directly sent to the $k^\text{th}$ decoder layer. We can see that with each subsequent (orthogonal) addition of a component, there is a noticeable performance improvement. This proves that each of our components contributes sizeably to the final performance.

\begin{table}[tbp]
\small
\centering
\ra{1.2}
\caption{Ablation of different models.}
\label{table:4}
\begin{tabular}{c c c r r}
\toprule
ResBlock & Augmentation & AB & F1 & AUC\\
\midrule
\xmark & \xmark & \xmark & 0.8202 & 0.9793 \\
\cmark & \xmark & \xmark & 0.8250 & 0.9806 \\
\cmark & \cmark & \xmark & 0.8275 & 0.9828 \\
\cmark & \cmark & Type-1 & 0.8322 & 0.9809 \\
\cmark & \cmark & Type-2 & 0.8407 & 0.9833 \\
\bottomrule
\end{tabular}
\end{table}

\section{Conclusions}

\label{sec:conclude}

In this paper we present a novel architecture Attention W-Net, which is focuses on improving gradient flow in U-Net based architectures. We achieve the same in two ways, firstly by improving skip connections by attending over the encoder signal using the decoder features so that them being summed up results in a more meaningful signal as the decoder branch restructures the high dimensional features expanded by the encoder branch into segmentation maps. Secondly, we incorporate a modified Residual Block and use careful image augmentation to add a meaningful regularisation signal to our training process. The Residual Block helps in preventing overfitting by using Shared Weights, dropout and BatchNorm layers, whereas the image augmentation ameliorates the issues arising from having less data by making the model more robust to distortions by injecting the same stochastically into the training process. Our model has a sizeable performance gain over the backbone LadderNet, proving the utility of our novelty, and showcases top performance among the latest architectures as well.

\bibliographystyle{IEEEtran}
\bibliography{latex12}

\end{document}